\documentclass[aps,prl,twocolumn]{revtex4-1}
\usepackage{amssymb}
\usepackage{amsmath,bm}
\usepackage{graphicx,color}
\usepackage{bbm}
\usepackage{multirow}
\newcommand{\B}[1]{{\bm{#1}}}

\begin{document}

\title{Scaling theory of shear-induced inhomogeneous dilation in granular matter}

\author{Prasenjit Das$^1$, H. George E. Hentschel$^{1,2}$ and Itamar Procaccia$^1$ }
\affiliation{$^1$Department of Chemical Physics, The Weizmann Institute of Science, Rehovot 76100, Israel.\\ $^2$ Department of Physics, Emory University, Atlanta, Georgia 30322,USA.}

\begin{abstract}
Shearing with a finite shear rate a compressed granular system results in a region of grains flowing over a compact, static assembly. Perforce this region is dilated to a degree that depends on the shear rate, the loading pressure, gravity, various material parameters, and the preparation protocol. In spite of numerous studies of granular flows a predictive theory of the amount of dilation is still lacking. Here, we offer a scaling theory that is focused on such a prediction as a function of shear rate and the dissipative parameters of the granular assembly. The resulting scaling laws are universal with respect to changing the interparticle force laws.
\end{abstract}

\maketitle
The least understood regime of granular flow is the ``intermediate regime" between the solid and the kinetic regimes~\cite{08FP,04MiDi,05daCruz,06JFP,08FP,08PR,10DD}. In this regime grains  interact  both through enduring contacts and through collisions~\cite{12ACLAE}. The flow characteristics depend on many variables and also on preparation protocols, making it hard to develop predictive theories to foretell how a compressed granular system would respond to finite shear rates~\cite{96JNB}.

In this Rapid Communication we focus on the inhomogeneous dilation that results from shearing a compressed binary assembly of circular disks~\cite{02Lem,11Sir}. It was already pointed out almost 30 years ago by Thompson and Grest~\cite{91TG} that understanding the dependence of the amount of dilation as a function of the experimental and material parameters is hard. In their words ``analytical treatments are difficult because the boundary conditions and velocity distribution functions are poorly understood, and microstructure induces complex correlations among grains~\cite{90Cam,87JJ,90JLNW,03SLG}." In spite of the large literature on dense granular flows since~\cite{91TG}, these words are still relevant.

\begin{figure}
\includegraphics[scale=0.80]{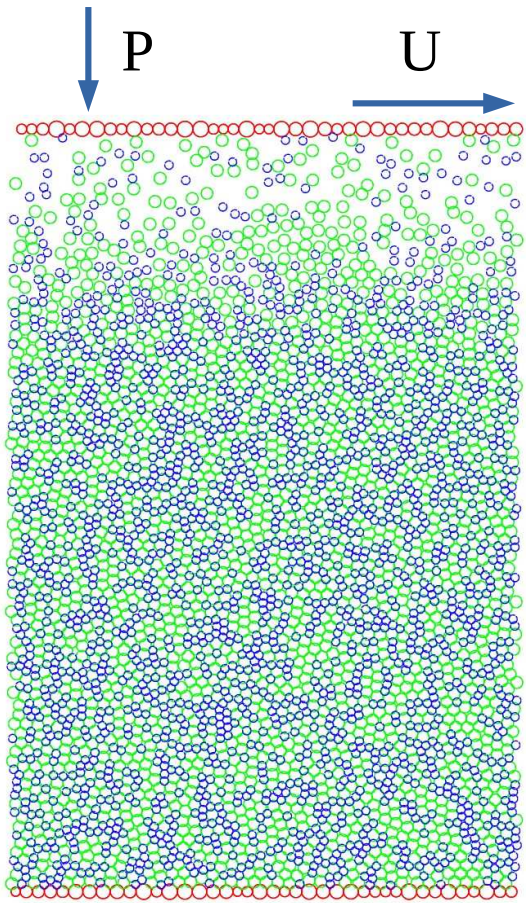}
\caption{An example of the simulational setup with the dilated layer that is the subject of this Rapid Communication.}
\label{diagram}
\end{figure}

In this Rapid Communication we propose that some of these difficulties can be surmounted by employing scaling concepts. To do so one needs to single out the important parameters that can be incorporated into a scaling theory without leaving out any essential physics that may mar the predictability of such a theory. As did Thompson and Grest we will consider here a two-dimensional assembly of disks confined in the $\hat y$ direction by two walls, with periodic boundary condition in the horizontal $\hat x$ direction. Contrary to Ref.~\cite{91TG} we use a binary mixture of 4000 disks of mass $m=1$, half with diameter $\sigma_1=2R_1=1$ and the other half with $\sigma_2=2R_2=1.4$. Preparing binary mixtures of disks requires some care to avoid segregation. We have prepared our system to shearing in the following way: First we select $N$ random positions for the $N$ disks in a box of the final width but a height large enough to avoid any initial overlap, having zero pressure. Next, the upper piston is pushed quasistatically in the absence of gravity to reduce the box height. This continues until we achieve a small finite pressure. At this point gravity is turned on and the piston is subjected to a wanted pressure, and the system is allowed to equilibrate employing a global viscous damping. After equilibration the system is ready for shearing.  The boundary walls are made of glued disks with random diameters in the range of $[\sigma_1,2\sigma_1]$. The upper wall has a mass $M=100$ and is free to move. The lower wall is fixed at $y=0$.  Gravity with $g=1$ is applied, such that the unit of length is $\sigma_1$ and the unit of time is $\sqrt{\sigma_1/g}$. To test our theory we perform numerical simulations in which the upper wall is pressed down with a pressure $P$ and pulled in the $\hat x$ direction at a fixed velocity $U\hat x$. During shearing we update positions and velocities of the disks using the velocity-Verlet algorithm with a time step $\Delta t=0.0001$. The primary observed response is the dilation $\delta$ that develops spontaneously under nonequilibrium steady state conditions (see Fig.~\ref{diagram}).

The numerical simulations are standard~\cite{79CS,01SEGHLP}. When the disks are compressed they interact via a normal force.  Particles $i$ and $j$, at positions ${\B r_i, \B r_j}$ with velocities ${\B v_i, \B v_j}$ experience a relative normal compression on contact given by $\Delta_{ij}= r_{ij}-D_{ij}$, where $\B r_{ij}$ is the vector joining the centers of mass and $D_{ij}=R_i+R_j$; this gives rise to a  normal force $ \B F^{(n)}_{ij} $. The normal force is modeled either as a Hertzian contact, or a Hookean one.  Defining $R_{ij}^{-1}\equiv R_i^{-1}+R_j^{-1}$, the force magnitudes are,
\begin{eqnarray}
&&\B F^{(n)}_{ij}\!=\!k\Delta_{ij} \B n_{ij}-\frac{\gamma}{2} \B {v}_{n_{ij}},\nonumber\\
&&k = k'\sqrt{ \Delta_{ij} R_{ij}}, \ \gamma = \gamma'  \sqrt{ \Delta_{ij} R_{ij}},\  \quad \text{Hertzian},\nonumber\\
&&k = k' \ ,\ \gamma = \gamma' \ ,  \quad \text{Hookean}. \
\label{forces}
\end{eqnarray}
Here, $\B n_{ij}$ is the normal unit vector, $k'=2\times 10^5$ is the spring stiffness. $\gamma'$ is a viscoelastic damping constant that is a variable parameter in our simulations. $\B {v}_{n_{ij}}$ is the normal component of the relative velocity between two particles given by $\B {v}_{n_{ij}}= (\B {v}_{ij}\cdot\B n_{ij})\B n_{ij}$, where $\B {v}_{ij} = \B {v}_{i} - \B {v}_{j}$. The translational acceleration of particles is calculated from Newton's second law; the total force on particle $i$ is given by $\B F^{\rm (tot)}_{i}= \sum_{j}\B F^{(n)}_{ij}$.

\begin{figure}
\includegraphics[scale=0.34]{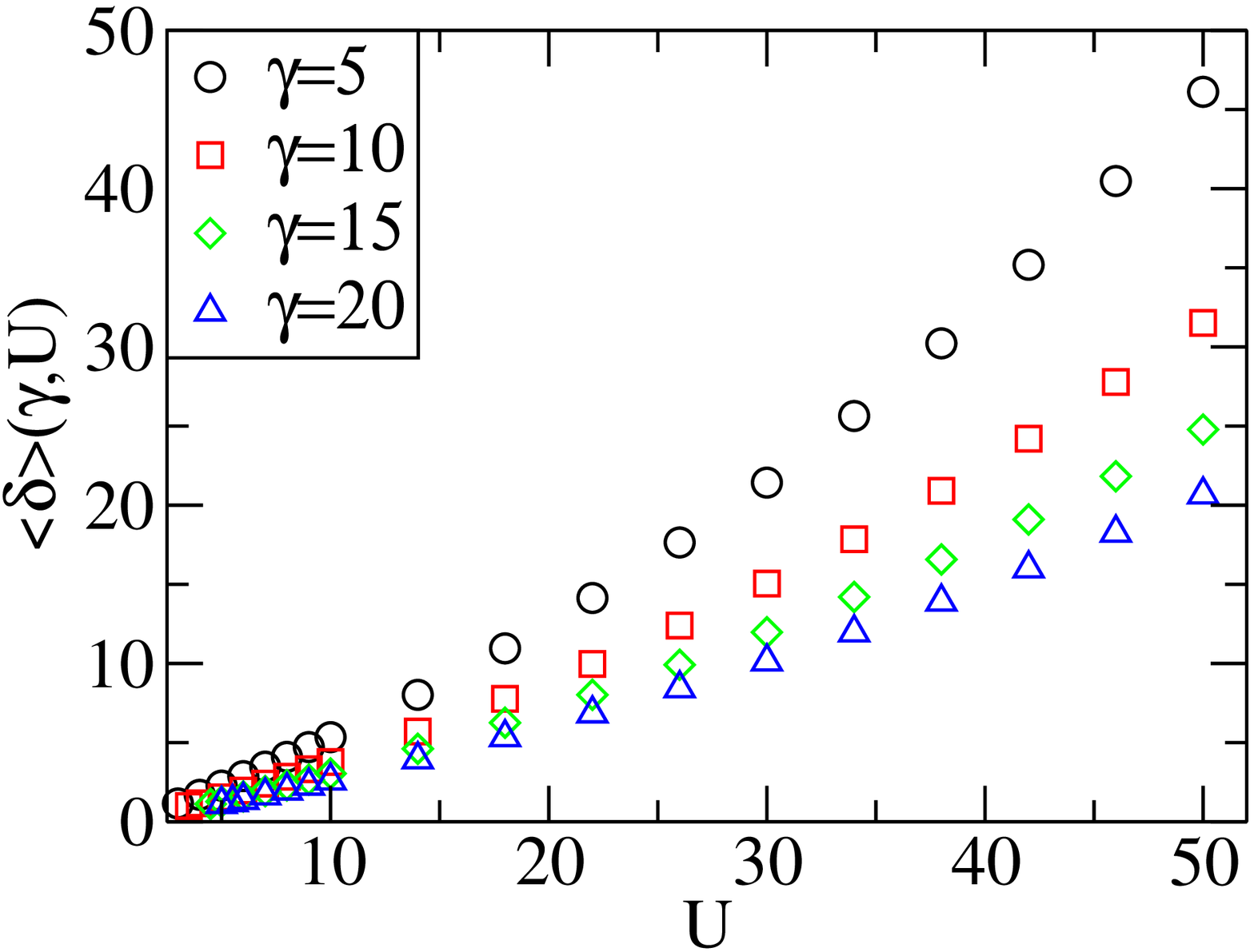}
\vskip 0.3 cm
\includegraphics[scale=0.35]{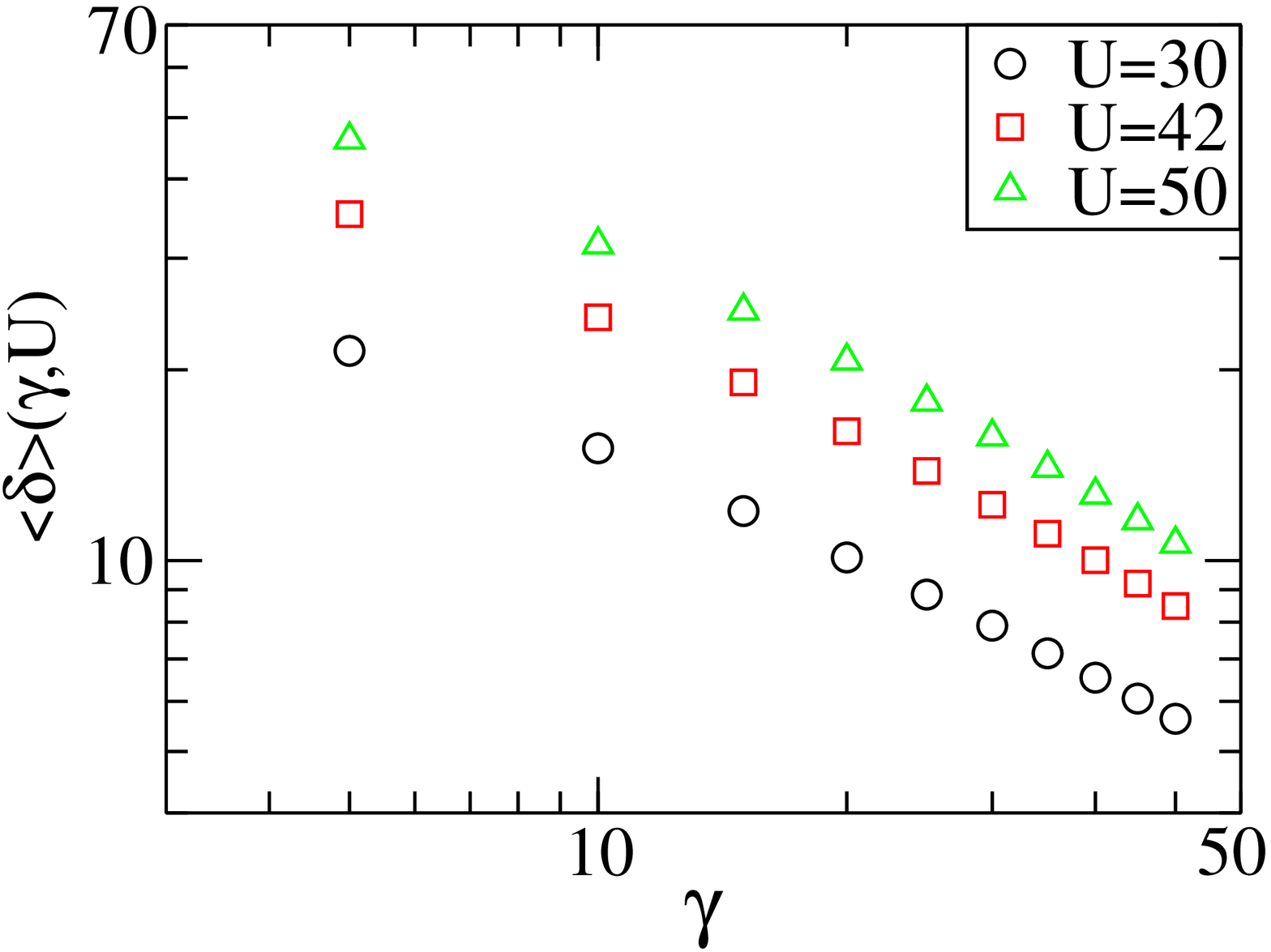}
\caption{Upper panel: Typical results for the dependence of $\langle \delta\rangle (\gamma,U)$ on $U$ for four different values of the damping coefficient $\gamma$. Lower panel: Typical results for the dependence of $\langle \delta\rangle (\gamma,U)$ on the damping coefficient $\gamma$ for three different values of the piston velocity $U$.}
\label{del-U-gam}
\end{figure}

\begin{figure}
\includegraphics[scale=0.35]{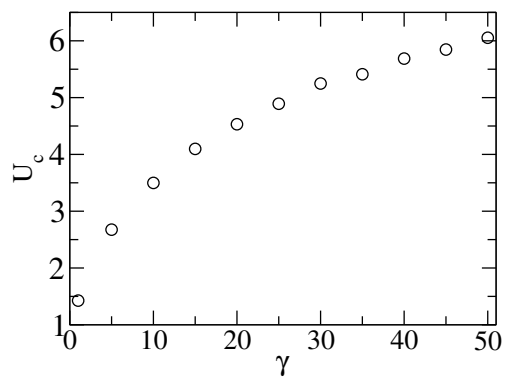}
\caption{$U_c$ for different values of $\gamma$.}
\label{Uc}
\end{figure}

The numerical simulations and the studied dilation depend on many parameters, namely, $P, g, U, m, M, R_1, R_2, k, \gamma$, the binary force laws, the grain shapes~\cite{14Wegner}, and the boundary conditions. In the present simulations we fix all the parameters with $P=10mg/\sigma_1$ and the $x$ dimension of the box $L_x=60\sigma_1$. First, we discuss Hookean binary forces, and at the end we show that Hertzian contacts lead to the same scaling theory. We study the amount of dilation as a function of changing $U$ and $\gamma$. Denoting the rest height of the box by $L_y(0)$ we measure the actual height $L_y(t)$ of the upper wall which is a function of time and the $U$. Once the flow is reaching a steady state we denote the dilation as $\delta(t,\gamma,U)\equiv L_y(t) - L_y(0)$ and its average as
\begin{equation}
\langle \delta\rangle (\gamma,U) \equiv \lim_{\tau\to \infty}\frac{1}{\tau} \int_T^{T+\tau} \!\!\!\!\!\delta(t,\gamma,U) dt
\end{equation}
The data that we want to understand are presented in Fig.~\ref{del-U-gam}, showing how the dilation depends on $U$ for different values of $\gamma$ and vice versa.
It is also interesting to note that for any given $\gamma$ there is a minimal value of $U$ below which no dilation is observed. To find this value of $U$, denoted as $U_c(\gamma)$, we extrapolate the data of $\delta$ vs $\gamma$ to $\delta=\sigma_1$ and read the value of $U_c$. A plot of $U_c$ vs $\gamma$ is shown in Fig.~\ref{Uc}. Note that in our theory we only consider dilations larger than a few disk diameters.

\begin{figure}
\includegraphics[scale=0.35]{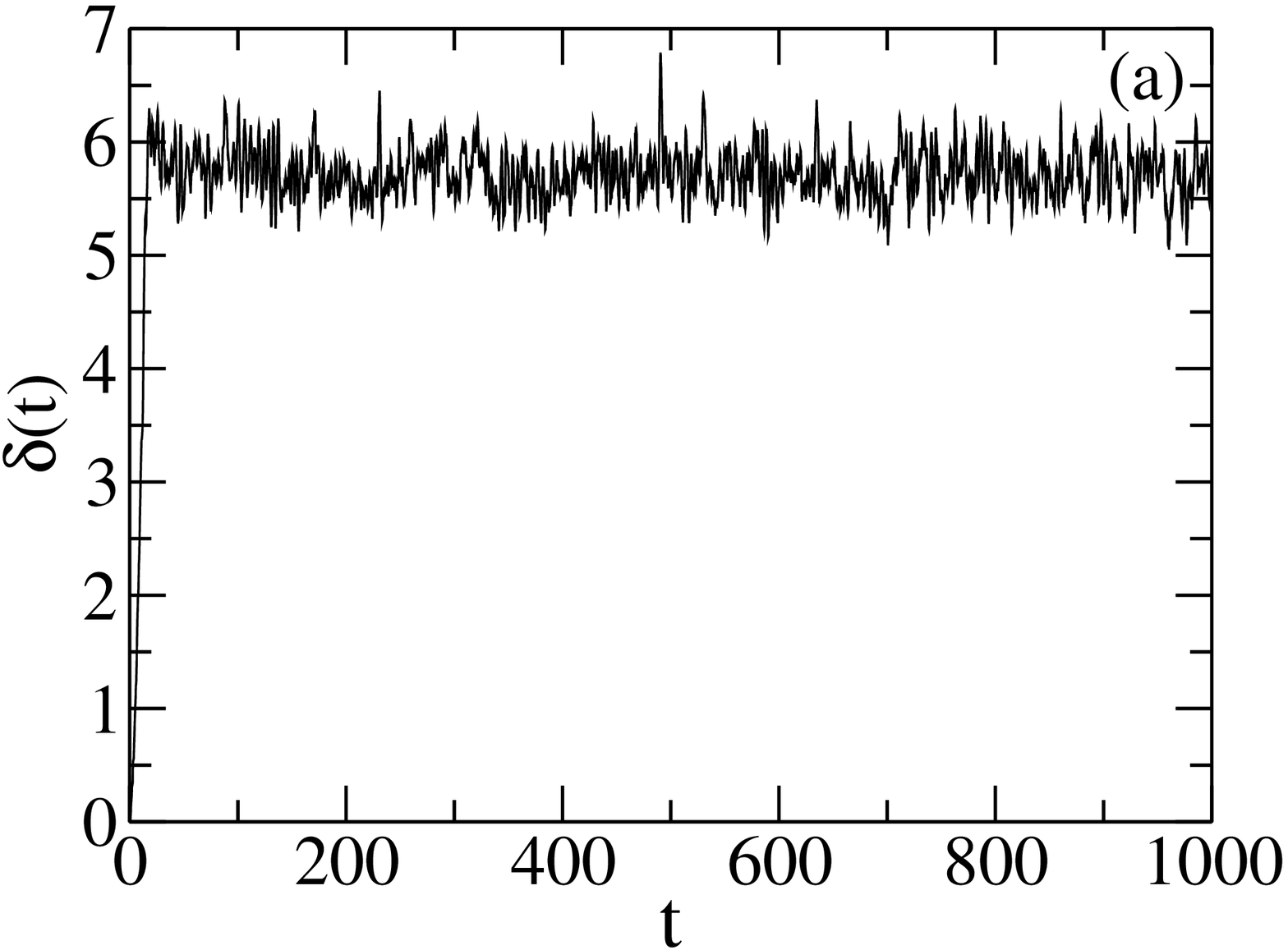}
\vskip 0.3cm
\includegraphics[scale=0.35]{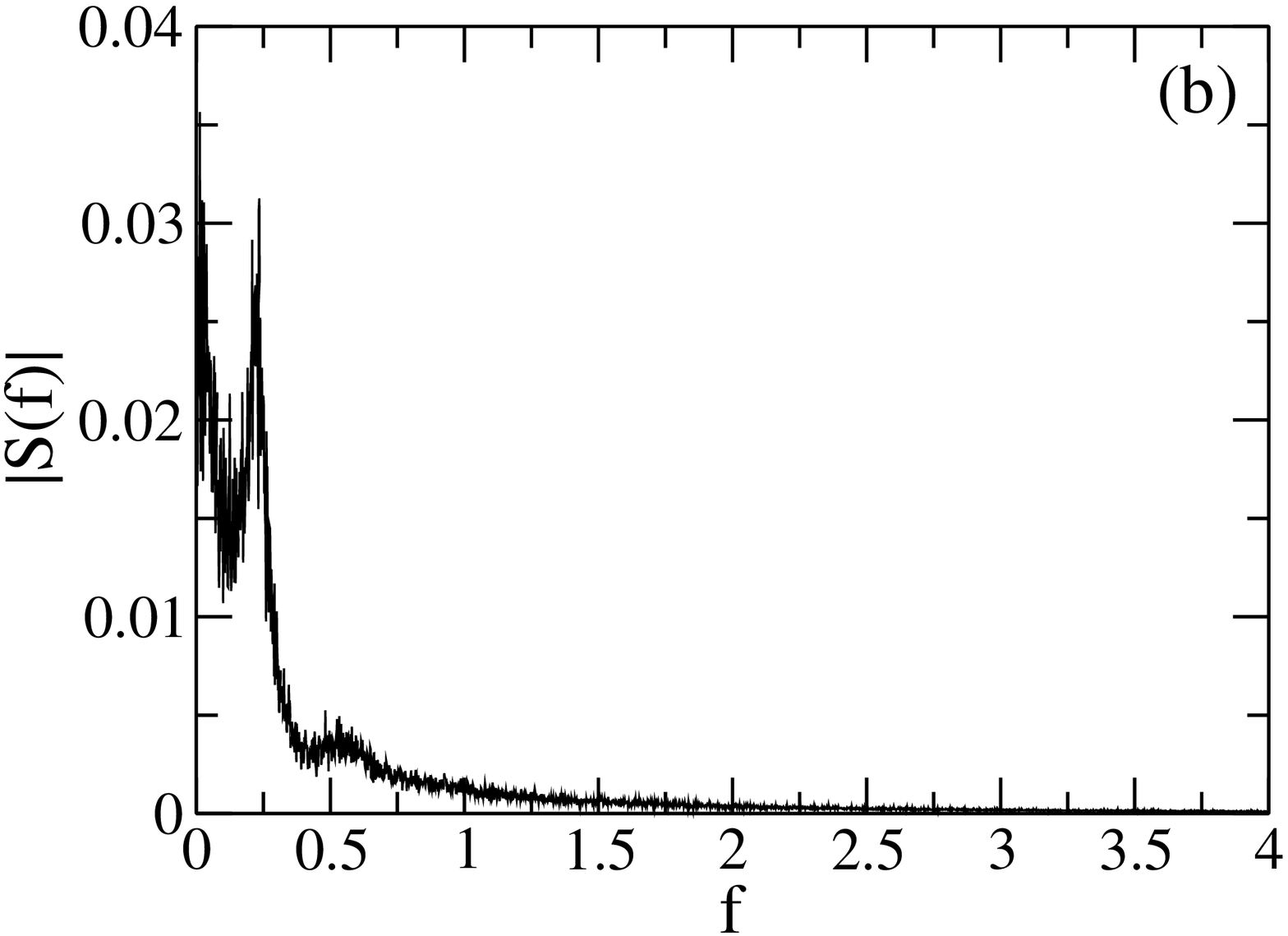}
\caption{Upper panel: A typical temporal trajectory of the dilation $\delta(t)$. Lower panel: A power spectrum of the above trajectory. In both panels $U=14$ and $\gamma=10$. The peak frequency is determined by the time that sound waves travel the system hight.}
\label{traj}
\end{figure}

\begin{figure}
\includegraphics[scale=0.35]{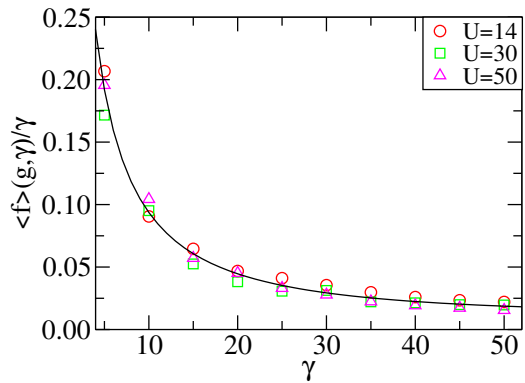}
\caption{A plot of the scaling function $F(\sqrt{g/\sigma}/\gamma)$. The important lesson is that it is independent of $U$. Shown is a data collapse for three values of $U$. The line is the analytic form Eq.~(\ref{scfun}).}
\label{scaling1}
\end{figure}

As said above, we want to approach the problem with the help of a scaling theory. The first step is to understand the typical time scale in this problem. Observing a generic trajectory of $\delta(t,\gamma,U)$ as shown in the upper panel of Fig.~\ref{traj}, we see that it follows a noisy periodic behavior. This is underlined by the spectrum shown in the lower panel of the same figure. Although there exists a prominent peak in the spectrum, it remains broad enough, and therefore we cannot take the typical time scale as the inverse frequency of the peak.  Thus we define an average frequency which will set the time scale of granular collisions
\begin{equation}
\label{avf}
\langle f\rangle = \int_0^\infty f S(f) df/\int_0^\infty S(f) df \ .
\end{equation}
This frequency could  in principle depend on $U$, the gravitational acceleration $g$ and $\gamma$. A crucial simplifying feature of the scaling theory is that  $\langle f\rangle (U,g,\gamma)$ {\em does not} depend on $U$. We shall therefore make the scaling ansatz
\begin{equation}
\label{avf2}
\langle f\rangle (g,\gamma) = \gamma F(\sqrt{g/\sigma}/\gamma) .
\end{equation}
To test this ansatz we present the proposed scaling function $F(\sqrt{g/\sigma}/\gamma)$ as computed from simulations. Its lack of dependence on $U$ is clearly evident in Fig.~\ref{scaling1}.

The essence of the scaling theory is the identification of the balance between the rate of work done $\dot W$ by the piston that is moving at a fixed velocity $U$ and the rate of dissipation by the various dissipative mechanisms that maintain a steady state motion. The assumption is that the rate of work done is dominated by the interaction of the piston with the upper most layer of grains that are being sheared by the motion. Since there are about $L_x/\sigma$ grains interacting with the piston which imparts momentum of the order of $mU$ per unit time $\sigma/U$, we estimate the rate of work done as
\begin{equation}
\dot W \sim \frac{L_x}{\sigma}\frac{mU}{\sigma/U} U =\frac{L_xmU^3}{\sigma^2} \ .
\label{rate}
\end{equation}
This estimate indicates that for some velocity $U < U_c$ the rate of work done by the piston is so small that it dissipates without creating of a sizeable dilated layer. Below we will estimate the size of this critical  velocity.

The rate of dissipation is dominated by at least two processes. We denote by $\dot D_1$ the rate of viscous dissipation due to the intergranular force $m\gamma {\vec v}_{n_{ij}}$ between all pairs of particles $i$ and $j$ in the dilated region of size $\delta$. The number of particles in the dilated region scales as $\sim L_x\delta \phi /\sigma^2$ where $\phi$ is the average area fraction in the dilated region. The rate of work done on each of these particles is estimated as $m \gamma v_{ij}^2$. We also estimate $ v_{ij}\sim v_x(y+\sigma)-v_x(y) \approx (U \sigma/\delta )$. In total
\begin{equation}
\label{Wvis}
 \dot D_1 \sim  \frac{ L_x\delta \phi }{\sigma^2} m \gamma v_{ij}^2  \sim  \frac{ L_x\delta \phi }{\sigma^2} m \gamma (U \sigma/\delta )^2
\sim  L_x \phi m \gamma U^2/\delta.
\end{equation}
The other contribution to the rate of dissipation, denoted as $\dot D_2$, results from grain collisions that occur at the bottom of the dilated layer as they strike the compact granular medium. These collisions result in the  piston fluctuations shown in Fig.~\ref{traj}. We thus
estimate:
\begin{equation}
\label{Wcoll}
\dot D_2  \sim (L_x/\sigma) m {\ddot \delta} {\dot \delta} ,
\end{equation}
where $\ddot \delta$ is the typical magnitude of the grain acceleration, while $\dot \delta$ is the typical grain speed at the bottom of the dilated region. As before $(L_x/\sigma)$ is the number of grains involved. At this point employ the scaling ansatz Eq.~(\ref{avf2}) for the average frequency to estimate the average velocities and accelerations of grains near the bottom of the dilated region
\begin{eqnarray}
\label{eqn3}
{\dot \delta} &\sim & \langle f\rangle \delta \approx \gamma F(\sqrt{g/\sigma}/\gamma) \delta  \nonumber \\
{\ddot \delta} & \sim  &\langle f\rangle^2 \delta \approx \gamma^2 F^2(\sqrt{g/\sigma}/\gamma) \delta .
\end{eqnarray}
Thus our final estimate for the collisional contribution to the rate of dissipation is
\begin{equation}
\label{Wcoll2}
\dot D_2  \sim (L/\sigma) m \langle f\rangle^3 \delta^2  \approx (L/\sigma) m \gamma^3 F^3(\sqrt{g/\sigma}/\gamma) \delta^2 \ .
\end{equation}
Balancing the estimates given by Eq.~(\ref{rate}) with Eqs.~(\ref{Wvis}) and (\ref{Wcoll2})  yields the algebraic equation for $\delta$
\begin{equation}
\label{Weq}
\delta^3 + a  U^2 = b U^3 \delta ,
\end{equation}
where
\begin{eqnarray}
\label{parameters}
a & =  & \frac{C_1 \phi \gamma \sigma}{\langle f\rangle^3} \approx  \frac{C_1 \phi  \sigma}{\gamma^2 F^3(\sqrt{g/\sigma}/\gamma)} \nonumber \\
b & =  &\frac{C_2}{\sigma \langle f\rangle^3}\approx  \frac{C_2}{\sigma \gamma^3 F^3(\sqrt{g/\sigma}/\gamma)} ,
\end{eqnarray}
and $C_1,C_2$ are dimensionless constants of $O(1)$.

\begin{figure}
\includegraphics[scale=0.35]{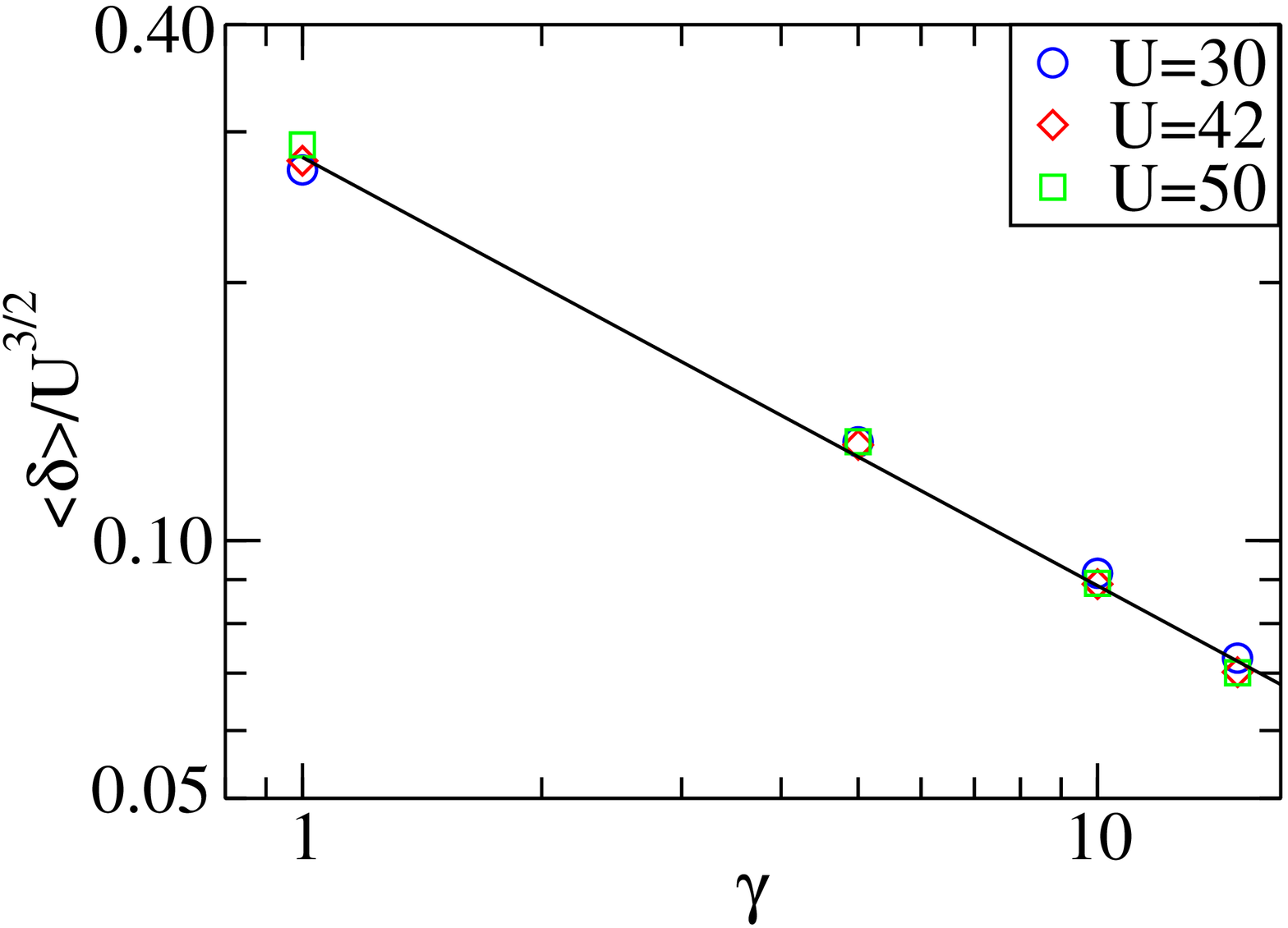}
\vskip 0.3 cm
\includegraphics[scale=0.35]{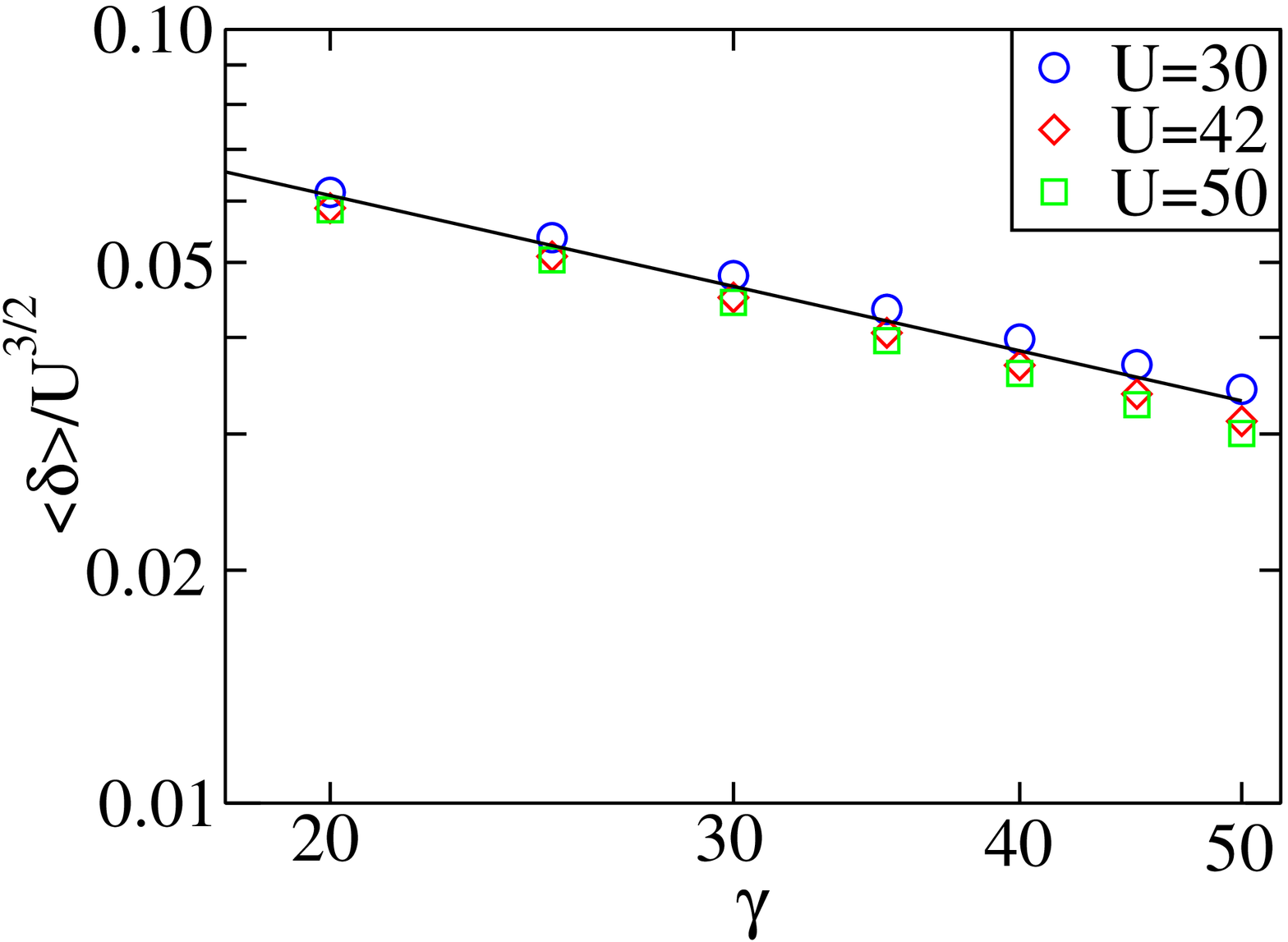}
\caption{Upper panel: Equation~(\ref{pre1}) presented in a log-log plot; the slope of the line is -1/2. Lower panel: The same test of the prediction Eq.~(\ref{pre2}); the slope is -2/3.}
\label{pred1}
\end{figure}

Consider first the solution of this equation for $U$ sufficiently large. Then the quadratic term in Eq.~(\ref{Weq}) is negligible and we find the solution
\begin{equation}
\delta(U,\gamma) = \sqrt{b}~ U^{3/2} \ .
\label{pred}
\end{equation}
We will show below that the dependence of $\delta$ on $U$ when $U$ is large agrees very well with this prediction. But the coefficient $\sqrt{b}$ contains a complex $\gamma$ dependence that comes from the scaling function $F(\sqrt{g/\sigma}/\gamma)$. This function does not have a simple scaling form in the whole regime of our simulations. The scaling function $F(x)$ followed from the definition of an average frequency Eq.~(\ref{avf}), and it depends on whether we have a thin dilation layer ($\gamma$ large) or thick dilation layer when $\gamma$ is small. Consider first the function $F(x)$ for large $x$, (or $\gamma$ small). We expect that in this limit the dilation would be large, and therefore insensitive to the grain size $\sigma$. Using Eq.~(\ref{pred}) we therefore demand that $\sqrt{b}$ would be independent of $\sigma$. This will occur if $F(x) \sim x^{2/3}$ in this regime. On the other hand for $x\to 0$ or $\gamma\to \infty$, we expect that the typical frequency will be determined fully by $\gamma$, so $F(x\to 0)=$ const. In the intermediate range of moderate values of $x$ the scaling function can have yet another form that we cannot determined by pure theory. We use therefore the numerical results as shown in Fig.~\ref{scaling1}, and find the best fit to the following function,
\begin{equation}
F(x) = 0.035+2.781 x^{2/3} -1.942x^{5/9}.
\label{scfun}
\end{equation}
The exponent $5/9$ was chosen as the closest rational number to the numerically obtained best fit
of an exponent 0.55$\pm 0.03$. This result controls the scaling form of $\delta(U,\gamma)$ for the whole available range of $\gamma$ and $U$.
For $\gamma$ relatively small (i.e., large $x$) $F(x)\sim x^{2/3}$. Then we predict
\begin{equation}
\delta(U,\gamma) \sim \frac{U^{3/2}}{\gamma^{1/2}},~~~~~~ \gamma \text {~small}.
\label{pre1}
\end{equation}
In the upper panel of Fig.~\ref{pred1} we display $\delta(U,\gamma)/ U^{3/2}$ for all values of $U\ge 30$ and observe that we have data collapse with the expected dependence of $\gamma^{-1/2}$. In the intermediate regime of larger values of $\gamma$ (but still sufficiently large $U$), $F(x) \sim x^{5/9}$, and we expect that
\begin{equation}
\delta(U,\gamma) \sim \frac{U^{3/2}}{\gamma^{2/3}}, ~~~~~~\gamma \text {~large}.
\label{pre2}
\end{equation}
The lower panel of Fig.~\ref{pred1} supports this prediction as well.

\begin{figure}
\vskip 0.3 cm
\includegraphics[scale=0.35]{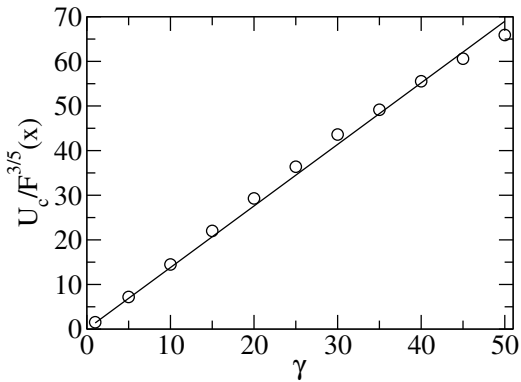}
\caption{A test of the theoretical prediction for $U_c$, Eq.~(\ref{ucrit}) .}
\label{FigUc}
\end{figure}

At this point we should study the dilation at {\em low} velocities, and understand the critical velocity $U_c$ shown in Fig.~\ref{Uc}. To this aim we solve the cubic equation Eq.~(\ref{Weq}) exactly. This equation has three roots, only one of which yields a real and positive value for $\delta$,
\begin{eqnarray}
\label{root}
\delta  &= &\frac{(2/3)^{1/3} b U^3}{[-9 a U^2 + \sqrt{81 a^2 U^4 - 12 b^3 U^9}]^{1/3}} \nonumber\\&+ &\frac{[-9 a U^2 + \sqrt{81 a^2 U^4 - 12 b^3 U^9}]^{1/3}}{18^{1/3}}.
\end{eqnarray}
This solution becomes singular when the square root in Eq.~(\ref{root}) goes through zero at some critical velocity $U= U_c(\gamma)$. This occurs at  $27 a^2 U_c^4 = 4 b^2 U_c^9$ or
\begin{equation}
\label{ucrit}
U_c = [ \frac{27 a^2}{4 b^3}]^{1/5}= [\frac{27 \phi^2}{4}]^{1/5} (\gamma \sigma ) F^{3/5}(x) \ .
\end{equation}
This result indicates that we should replot the data in Fig.~\ref{Uc} divided by $F^{3/5}(x)$ as a function of $\gamma$ to get a linear plot with slope of the order of unity. Figure~\ref{FigUc} supports this conclusion very well.

\begin{figure}
\vskip 0.3 cm
\includegraphics[scale=0.35]{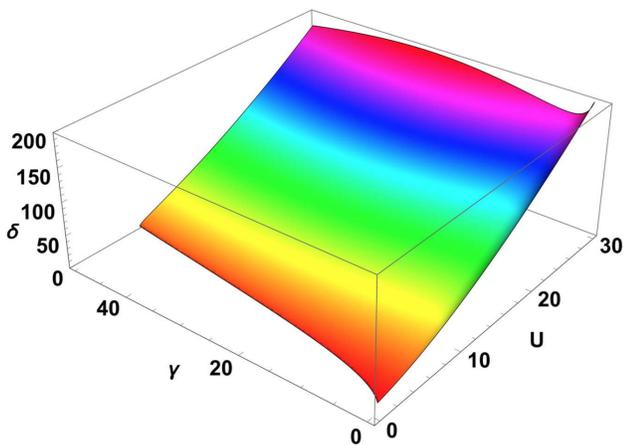}
\caption{A theoretically predicted phase diagram using the scaling function $F(x)$ from
Eq.~(\ref{scfun}).}
\label{final}
\end{figure}

Finally, we can summarize the predictions of the scaling theory as a phase diagram of $\delta$  as a function of $\gamma$ and $U$. To this aim we plot $\delta$ from Eq.~(\ref{root}) using  the expressions for $a$ and $b$ from Eq.~(\ref{parameters}) and the scaling function Eq.~(\ref{scfun}). The result is shown in Fig.~\ref{final}. One should notice the line of  $U_c(\gamma)$ which indicates the onset of dilation as a function of $U$ and $\gamma$. In addition the theory indicates a fast growing dilation for small $\gamma$ and large $U$.

\begin{figure}
\includegraphics[scale=0.25]{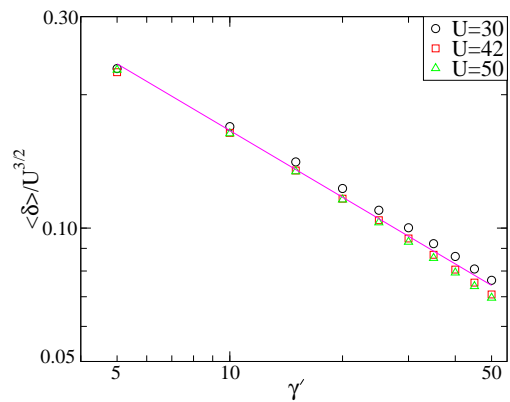}
\caption{The average dilation compensated by $U^{3/2}$ for different values of $\gamma'$. Here the binary forces are Hertzian and we find good agreement with Eq.~(\ref{pre2}). }
\label{scale}
\end{figure}

The reader should note that changing the interaction force from Hookean to Hertzian does not change the scaling theory. The only effect is renormalizing $\gamma$.
 \begin{figure}
 \vskip 0.3 cm
\includegraphics[scale=0.25]{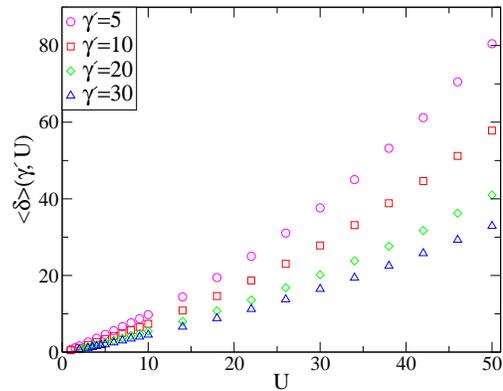}
\caption{Typical results for the dependence of $\langle \delta\rangle (\gamma,U)$ on $U$ for four different values of the damping coefficient $\gamma'$. Here the binary forces are Hertzian. }
\label{Hertz}
\end{figure}
It should be stressed that the scaling exponent $2/3$ in the scaling function was derived solely on the demand that for a thick dilation layer the amount of dilation should be independent of $\sigma$. On the other hand this determines the dependence of $\delta$ on $U$ when $\gamma$ is small in the form of Eq.~(\ref{pre1}). This result should be therefore universal, also independent of the interparticle binary force. To test this we have repeated our calculations with Hertzian rather than Hookean forces. The result is shown in Fig.~\ref{Hertz} which indeed supports the universality of Eq.~(\ref{pre1}). In Fig.~\ref{scale} we show that the average dilation compensated by $U^{3/2}$ agree with the scaling law Eq.~(\ref{pre1}). The straight line in Fig.~\ref{scale} has a slope -1/2.
The reader should note that in the Hertzian case $\gamma$ is effectively smaller than in the Hookean case [cf. Eq.~(\ref{forces})], and this is why the ``small $\gamma$" regime extends all the way to $\gamma'\approx 50$ in Fig.~\ref{scale}.

The use of scaling concepts in fluid flows has a long and honorable history. However, the identification of bifurcations and instabilities in granular flows has not been treated with scaling concepts nearly as often. The success of applying scaling concepts to polymer physics~\cite{79Gen}, with the coil-stretch transition under shear as an example, indicates that transitions of the type discussed here of dilation under shear should be subject to similar efforts. The theory described above is simple, assuming that all the pertinent dissipation mechanisms have been identified. Yet it appears to rationalize all the observed scaling laws and in addition it predicts the dependence of the threshold $U_c$ for all the simulated range of $U$ and $\gamma$. The onset of dilation itself is presently beyond the scope of the theory offered here. Approaching from the non-dilated side one expects a stick-and-slip regime with larger and larger fluctuations as the dilated layer forms. Clearly a rich field of research awaits a scaling approach.

\acknowledgments
This work had been supported in part by the US-Israel BSF and the ISF joint program with Singapore. We thank Oleg Gendelman for several useful discussion.


\end{document}